# Dephasing in Open Quantum Dots

A.G. Huibers, M. Switkes, and C.M. Marcus

*Department of Physics, Stanford University, Stanford, California 94305*

K. Campman and A.C. Gossard

*Materials Department, University of California, Santa Barbara,*

*Santa Barbara, California 93106*

Shape-averaged magnetoconductance (weak localization) is used for the first time to obtain the electron phase coherence time $\tau_\phi$ in open ballistic GaAs quantum dots. Values for $\tau_\phi$ in the range of temperature $T$ from 0.335 to 4 K are found to be independent of dot area, and are not consistent with the $\tau_\phi \propto T^{-2}$ behavior expected for isolated dots. Surprisingly, $\tau_\phi(T)$ agrees quantitatively with the predicted dephasing time for disordered two-dimensional electron systems.

Decoherence is the process by which the quantum mechanical properties of a microscopic system are transformed into the familiar classical behavior seen in macroscopic objects. Mesoscopic electronic systems, which exhibit strong coherent quantum mechanical effects such as weak localization and universal conductance fluctuations (UCF), are ideal for investigations of decoherence. The key quantity in these phenomena is the phase coherence time, $\tau_\phi$, which determines the energy and length scales at which quantum behavior is seen. Considerable theoretical [1-4] and experimental [5-9] study has been directed toward understanding the mechanisms responsible for the loss of phase coherence (dephasing) and their dependence on temperature, dimensionality, and disorder.

Most studies of dephasing in mesoscopic systems have focused on disordered 1D and 2D conductors, where the dimensional crossover for quantum corrections to transport and interactions responsible for dephasing occurs when the sample width exceeds the phase coherence length $\ell_\phi = \sqrt{D\tau_\phi}$ and thermal length $\ell_T = \sqrt{D\hbar/k_B T}$, respectively ($D$ is the diffusion constant) [1, 4]. At low temperatures electron-phonon scattering rates are small compared to electron-electron scattering rates [10] and two electron-electron scattering mechanisms dominate dephasing: a large-energy-transfer scattering mechanism, which causes dephasing with a rate $\tau_{ee}^{-1} \propto T^2$ [11] — in a 2D electron gas (2DEG) this rate is

$$\tau_{ee}^{-1} = \frac{\pi}{4}\frac{(k_B T)^2}{\hbar E_F}\ln\frac{E_F}{k_B T} \tag{1}$$

for $k_B T << E_F$ where $E_F$ is the Fermi energy [11] — and a small energy-transfer (*Nyquist*) scattering mechanism, which gives a rate $\tau_{\phi N}^{-1} \propto T^{2/(4-d)}$ where $d$ is the dimensionality of the system ($d$ =1, 2) [1]. In a disordered 2DEG the Nyquist dephasing rate is

$$\tau_{\phi N}^{-1} = \frac{k_B T}{2\pi\hbar}\frac{\lambda_F}{\ell_e}\ln\frac{\pi\ell_e}{\lambda_F}, \tag{2}$$

where $\lambda_F$ is the Fermi wavelength and $\ell_e$ is the elastic mean free path. The total dephasing rate due to electron-electron scattering is approximately the sum of these rates, $\tau_\phi^{-1} \approx \tau_{\phi N}^{-1} + \tau_{ee}^{-1}$ [6, 12]. Measurements of $\tau_\phi(T)$ in disordered 2D and 1D semiconductors [6] and 1D metals [5] based on weak localization find good agreement with these theoretical results down to ~0.1 K. In clean 2D systems ($\ell_e \sim \ell_\phi$) the dephasing rate is expected to coincide with $\tau_{ee}^{-1}$ from Eq. (1), without the Nyquist contribution, consistent with experiments in high-mobility 2DEG samples [7, 13]. In isolated quantum dots (0D systems), a dephasing rate $\tau_\varphi^{-1} \propto\sim T^2$ is expected for intermediate temperatures ($\ell_T > L$ but $kT >> \Delta$, where $\Delta = 2\pi\hbar^2/m^* A$ is the mean level spacing for a dot of area $A$) with a rate comparable to Eq. (1) for ballistic dots, $\ell_e > L$ [2, 3, 14]. To our knowledge, there has been no theoretical discussion of $\tau_\phi$ in *open* quantum dots despite previous experimental investigation [8, 9].

In this Letter, we use a novel method based on the 0D analog of weak localization to measure $\tau_\phi(T)$ in ballistic GaAs quantum dots with areas ranging from 0.4 to 4 $\mu m^2$ and single-mode point-contact leads. We find that $\tau_\phi$ is independent of dot area and, surprisingly, that $\tau_\phi(T)$ is not proportional to $T^{-2}$ but rather is in good quantitative agreement with the theory for diffusive 2D conductors discussed above, including both $T^{-1}$ and $T^{-2}$ contributions. These conclusions are checked with a comparison to $\tau_\phi(T)$ measured three other ways in the same dots.

Our primary technique for determining $\tau_\phi$ is based on the magnetic field dependence of the weak localization correction to quantum transport, and is similar to standard methods used in diffusive 1D and 2D samples. This method is applied to quantum dots for the first time here, having only become possible due to recent theoretical developments [15, 16] based on random matrix theory (RMT) [17]. For irregularly-shaped quantum dots with two leads each supporting *N* channels (or lateral modes), RMT yields a zero-temperature average conductance $\langle g \rangle$ equal to the resistors-in-series value $\left(e^2/h\right)N$ at $B \neq 0$ but reduced to $\left(e^2/h\right) 2N^2/(2N+1)$ at $B = 0$ due to phase-coherent backscattering (or weak localization) [18]. Dephasing suppresses this difference, $\delta g \equiv \langle g \rangle_{B\neq 0} - \langle g \rangle_{B=0}$, by limiting the time over which backscattered electrons may contribute to interference. To incorporate dephasing into a quantitative theory, a fictitious voltage probe, or "ϕ-lead" supporting $\gamma_\phi$ modes, where

$$\gamma_\phi = \frac{2\pi\hbar}{\tau_\phi \Delta} \qquad (3)$$

is appended to the dot [19]. The RMT for this three-lead dot (two real leads plus the $\phi$-lead) then yields a suppressed weak localization correction [15],

$$\delta g \approx \frac{e^2}{h}\left(\frac{N}{2N+\gamma_\phi}\right), \qquad (4)$$

that models the effect of dephasing. Note that $\gamma_\phi$ is proportional to dot area, so a larger dot will exhibit a smaller $\delta g$ for a given $\tau_\phi$.

The $\phi$-lead model was recently improved by Brouwer and Beenakker by distributing the phase breaking throughout the dot rather than concentrating it at the location of a single lead [16]. The resulting expression for $\delta g$ in terms of $\gamma_\phi$ differs significantly from Eq. (4) for the case N=1 (Fig. 2, inset), though nearly coincides with Eq. (4) for N>1. We note that both the $\phi$-lead model and its distributed extension [16] ignore the effects of Coulomb charging on $\delta g$, which may be important particularly at N=1 [20]. The consistency between measured values of $\tau_\phi(T)$ using different methods and dot sizes suggests that any field-dependent charging effects are probably not corrupting the present measurement significantly.

Measurements on four quantum dots (Fig. 3, insets) with areas of 0.4 $\mu m^2$ (two dots), 1.9 $\mu m^2$ and 4.0 $\mu m^2$ ($\Delta$ = 17.9, 3.8, and 1.8 $\mu$eV respectively) are reported. The dots are formed by gate depletion of a 2DEG located 160 nm below the surface of a $GaAs/Al_{0.3}Ga_{0.7}As$ heterostructure (sheet density $n = 1.8\times10^{11}\ cm^{-2}$, mobility $\mu = 1.2\times10^{6}\ cm^2/Vs$, Fermi wavelength $\lambda_F$ = 60 nm and Fermi energy $E_F$ = 6.3 meV). The elastic mean free path (~11 $\mu$m) is larger than all device sizes, so that transport is ballistic within the dots. The dots were measured in a $^3$He cryostat at temperatures ranging from 335 mK to 4 K using standard 4-wire lock-in techniques at 105 Hz with 0.5 nA bias current— small enough not to affect transport due to heating ($I_{bias} = 0.5$ and 1 nA give identical results at base temperature). At these temperatures, weak localization and UCF are comparable in magnitude, as seen in the gray traces of Fig. 1. By averaging over gate-voltage-controlled shape distortions, UCF is averaged away leaving only the weak localization correction. The measurement procedure is illustrated in Fig. 1. First, $\mathrm{V_{pc1}}$ and $\mathrm{V_{pc2}}$ are swept in a raster to find the plateau with $N=1$ channel in each lead (bracketed region lower left inset). While the leads are maintained at one channel each, the shape of the dot is distorted using $\mathrm{V_{shape1}}$ and $\mathrm{V_{shape2}}$, creating an effective ensemble of dots. The 47 green points on the conductance landscape in the lower right inset indicate the positions in $\left(\mathrm{V_{shape1}},\mathrm{V_{shape2}}\right)$-space of the samples used in the ensemble. Figure 1 shows $g(B)$ at four of these 47 points, along with the average $\langle g(B)\rangle$ of all 47 used to determine $\delta g$.

Figure 2 shows $\delta g$ at $N=1$ as a function of temperature for the four devices. Using $\gamma_\phi(\delta g)$ from Ref. [16], each point in Fig. 2 is converted to $\gamma_\phi$ and then, using Eq. (3), to $\tau_\phi$. The resulting $\tau_\phi(T)$ is shown in Fig. 3. While dots with different areas have different values of $\delta g$, $\tau_\phi$ appears to be independent of area. The high temperature roll-off of $\tau_\phi$ seen in Fig. 3 for larger devices results from a breakdown of the model [16] when $\ell_\phi = v_F\tau_\phi$ becomes of order $L$, so that nonergodic trajectories dominate coherent backscattering. The inequality $L > \ell_\phi$ holds throughout the measured range of temperatures, however $L \sim \ell_T = v_F\hbar/k_BT$ at 2.2 K, 0.97 K and 0.69 K for the 0.4 μm$^2$, 1.9 μm$^2$ and 4.0 μm$^2$ dots respectively. As seen in Fig. 3, the temperature dependence of $\tau_\phi$ for all four dots falls between $\tau_\phi \propto T^{-2}$ and $\tau_\phi \propto T^{-1}$. The data cannot be fit by $\tau_{ee}$ alone (dashed line in Fig. 3) but are well fit by the sum dephasing rates for *diffusive 2D systems*, Eqs. (1) and (2), (solid line in Fig. 3) with $\ell_e = 0.25\mu\mathrm{m}$, giving $\tau_\phi^{-1}[\mathrm{ns}^{-1}]$ = 10.9 (T[K]) + 6.1 (T[K])$^2$. No saturation of $\tau_\phi$ is observed. The spin-orbit scattering time is expected to exceed the measured $\tau_\phi$ by at least an order of magnitude over the temperature range studied [21]. Significant spin-orbit scattering would lead to a local maximum of $\langle g(B) \rangle$ at $B = 0$, which is not observed.

To check the results based on weak localization amplitude at $N = 1$ ($\delta g_{N=1}$), we compare to three other measurements of $\tau_\phi(T)$ in the same devices (Fig. 4). The first comparison is to $\tau_\phi(T)$ obtained from weak localization amplitude at $N = 2$ ($\delta g_{N=2}$), measured as above, and using Eqs. (3) and (4) to convert from $\delta g_{N=2}$ to $\tau_\phi$. The $\delta g_{N=2}$ and $\delta g_{N=1}$ results are consistent within experimental error as shown in Figs. 4(a) and 4(b) for the 0.4 μm$^2$ and 4.0 μm$^2$ dots.

The second comparison is to $\tau_\phi(T)$ extracted from power spectra of UCF, a method described previously in Ref. [8]. This method makes use of the fact that UCF measured as a function of B in open quantum dots has an exponential power spectrum,

$$S(f) = S(0)e^{-2\pi B_c f} \tag{5}$$

for $kT >> \Delta$ [22] ($f$ is magnetic frequency in cycles/mT) with a characteristic magnetic field $B_c$ that depends on the dephasing rate,

$$\left(B_c/\varphi_0\right)^2 = \kappa(2N + \gamma_\phi). \tag{6}$$

where $\kappa$ is a geometry-dependent constant and $\varphi_0 = h/e$ is the quantum of flux [8]. Figure 4(c) shows power spectra of $g(B)$ for the 4.0 μm$^2$ dot, consistent with Eq. (5) over three orders of magnitude. A two-parameter fit of Eq. (5) to power spectra at each temperature gives $B_c(T)$ which yields $\tau_\phi(T)$ via Eq. (6), with $\kappa$ chosen as a best fit to the $\delta g_{N=1}$ data. Figure 4(d) compares $\tau_\phi(T)$ determined from UCF power spectra with that from $\delta g_{N=1}$, showing good agreement over the whole temperature range.

The final comparison is to $\tau_{\phi}(T)$ extracted from the *width* of the Lorentzian dip in average conductance around B=0 [23],

$$\langle g(B)\rangle = \langle g\rangle_{B\neq 0} - \frac{\delta g}{1+\left(2B/B_c\right)^2} \qquad (7)$$

Figure 4(e) shows traces of shape-averaged $\langle g(B)\rangle$ for the 4.0 μm$^2$ dot, along with two-parameter ($\delta g$ and $B_c$) fits to Eq. (7). Values for $\tau_{\phi}(T)$ in Fig. 4(f) are extracted from $B_c(T)$ using Eq. (6) with $\kappa$ chosen to give a best fit to the $\delta g_{N=1}$ results. It is noteworthy that several very different methods of determining $\tau_{\phi}$ agree within experimental error.

In summary, we have measured phase coherence times in open ballistic quantum dots using a new weak-localization method, as well as two other methods for comparison. We find: (1) consistency between the methods, (2) values for $\tau_{\phi}(T)$ for that do not depend dot area, and (3) an unexpected agreement between the experimental $\tau_{\phi}(T)$ and the theoretical prediction for a disordered 2D system with effective mean free path on the order of the device size and inconsistency with the $\tau_{\phi}(T) \propto T^{-2}$ expected for isolated dots. In particular, $\tau_{\phi}(T)$ appears to have significant contributions proportional to both $T^{-1}$ and $T^{-2}$, suggesting that perhaps some Nyquist-type dephasing mechanism is effective in open dots.

We thank Igor Aleiner, Boris Altshuler, Piet Brouwer, and Konstantin Matveev for very useful discussions. We gratefully acknowledge support at Stanford from the Office of Naval Research YIP program under Grant N00014-94-1-0622, the Army Research Office under Grant DAAH04-95-1-0331, the NSF-NYI program, the A. P. Sloan Foundation, and support for AGH from the John and Fannie Hertz Foundation. We also acknowledge support at UCSB by the AFOSR under Grant F49620-94-1-0158.

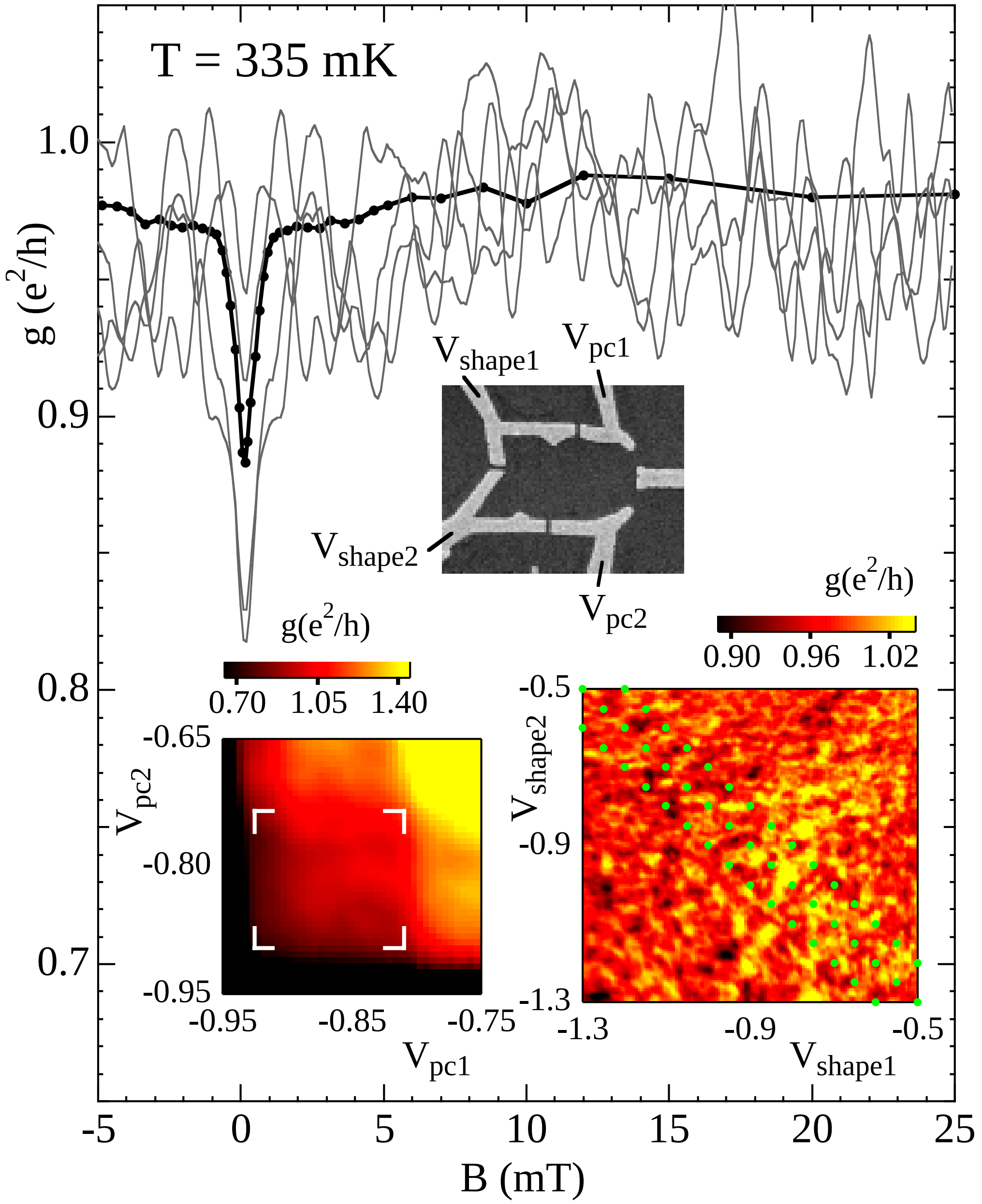

**Fig. 1.** Shape-averaged magnetoconductance (black) and four unaveraged conductance curves (gray) for the 4.0 μm$^2$ quantum dot (inset). Lower left inset: conductance as a function of $V_{pc1}$ and $V_{pc2}$ showing (bracketed) plateau with $N = 1$ channel in each point contact. Lower right inset: Conductance through dot as a function of $V_{shape1}$ and $V_{shape2}$ with green circles marking the 47 points at which magnetoconductance was measured.

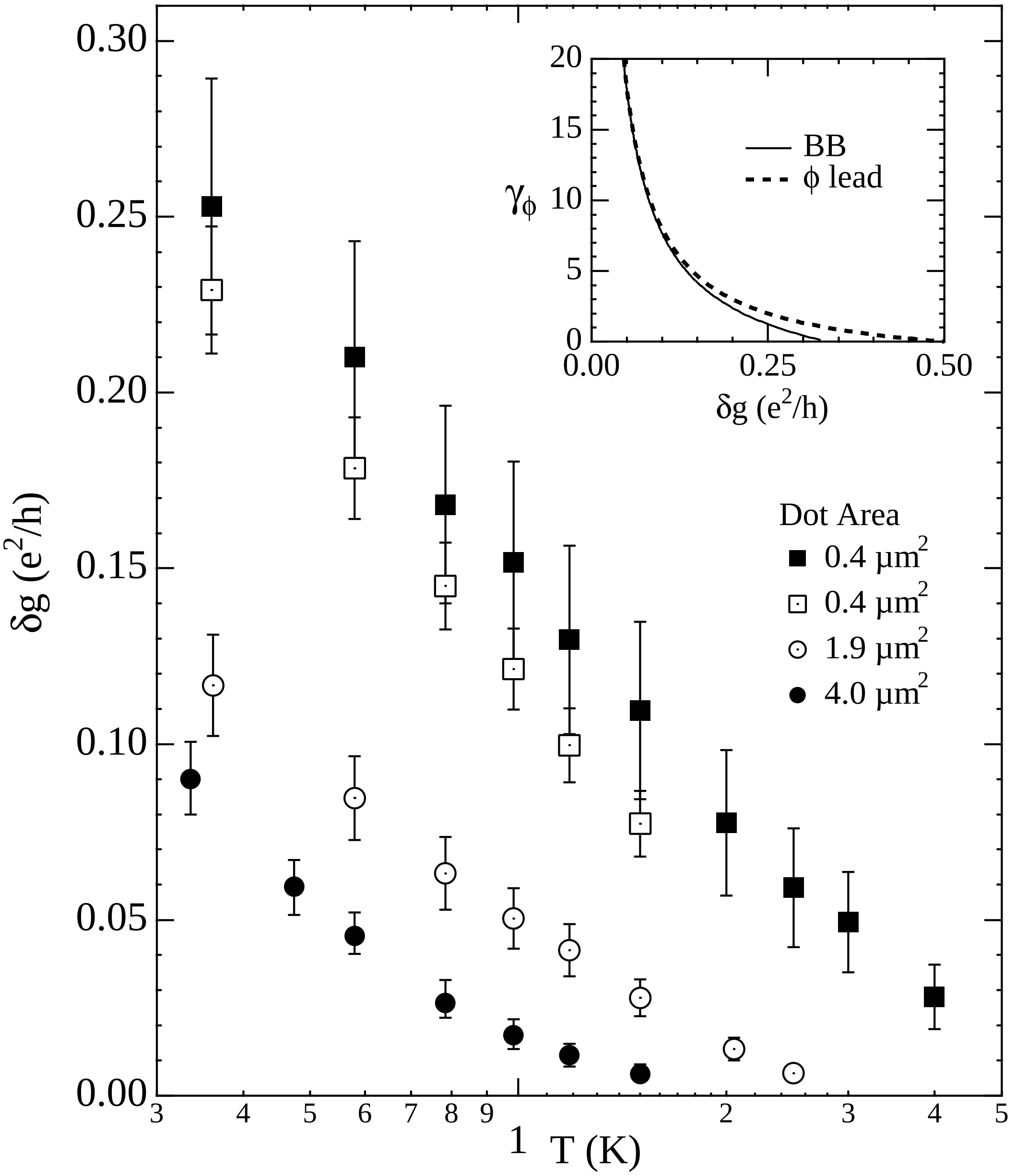

**Fig. 2.** Shape-averaged weak localization amplitude $\delta g$ vs. temperature $T$ for the four measured devices. Error bars reflect uncertainty in $\delta g$ as a result of conductance fluctuations remaining due to limited ensemble size. Inset: Theoretical phase breaking rate $\gamma_\phi(\delta g)$ using ϕ-lead model [15] (Eq. 3) and distributed voltage probe model (BB) [16].

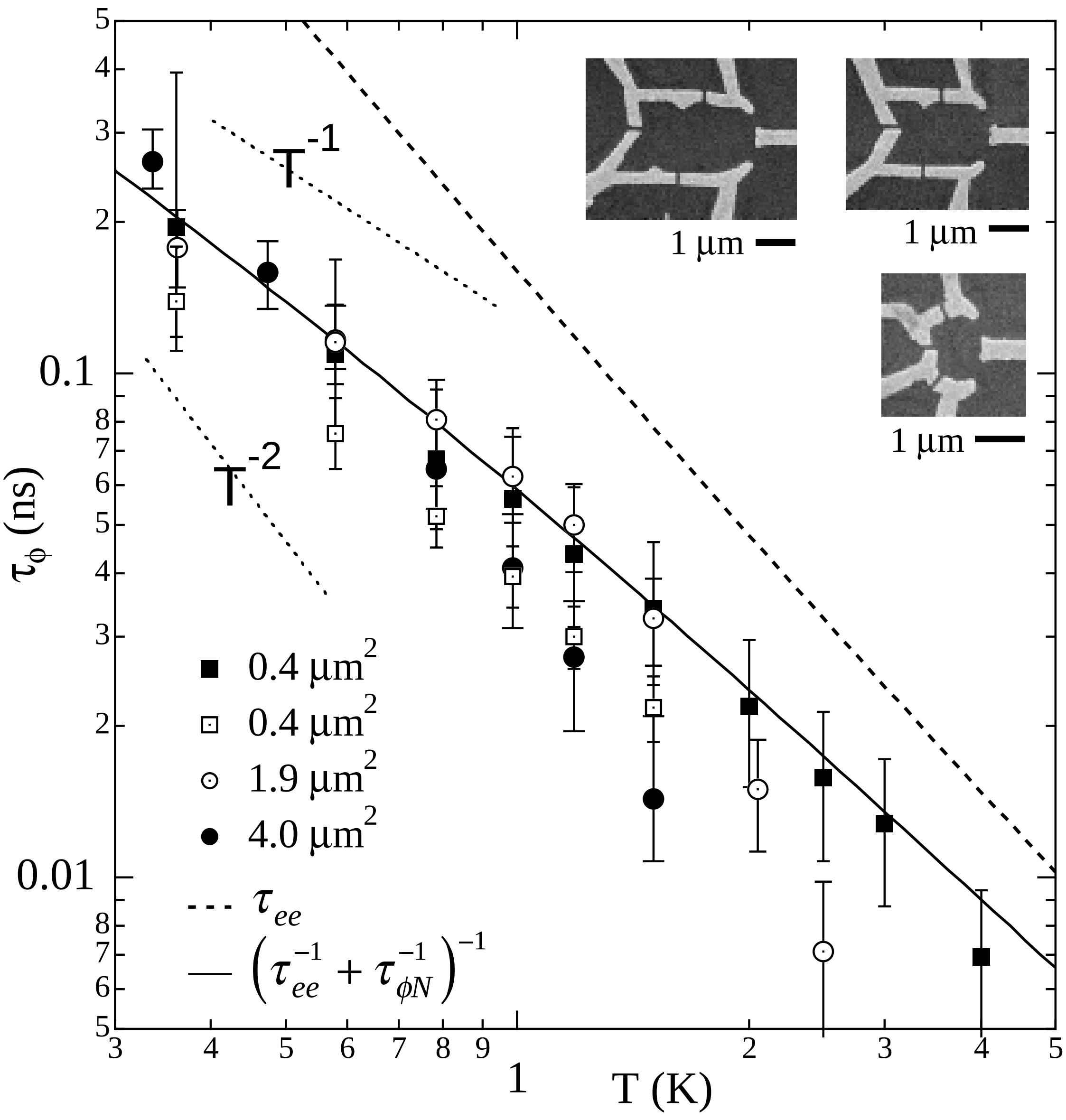

**Fig. 3.** Phase coherence time $\tau_\phi$ determined from $N$=1 weak localization. $\tau_{ee}$ from Eq. (2) (dashed), and $\tau_\phi^{-1} = \tau_{\phi N}^{-1} + \tau_{ee}^{-1}$ for a 2D disordered system with $\ell_e$ = 0.25 μm (solid) shown for comparison. Inset: micrographs of 4.0 μm$^2$, 1.9 μm$^2$, and 0.4 μm$^2$ dots.

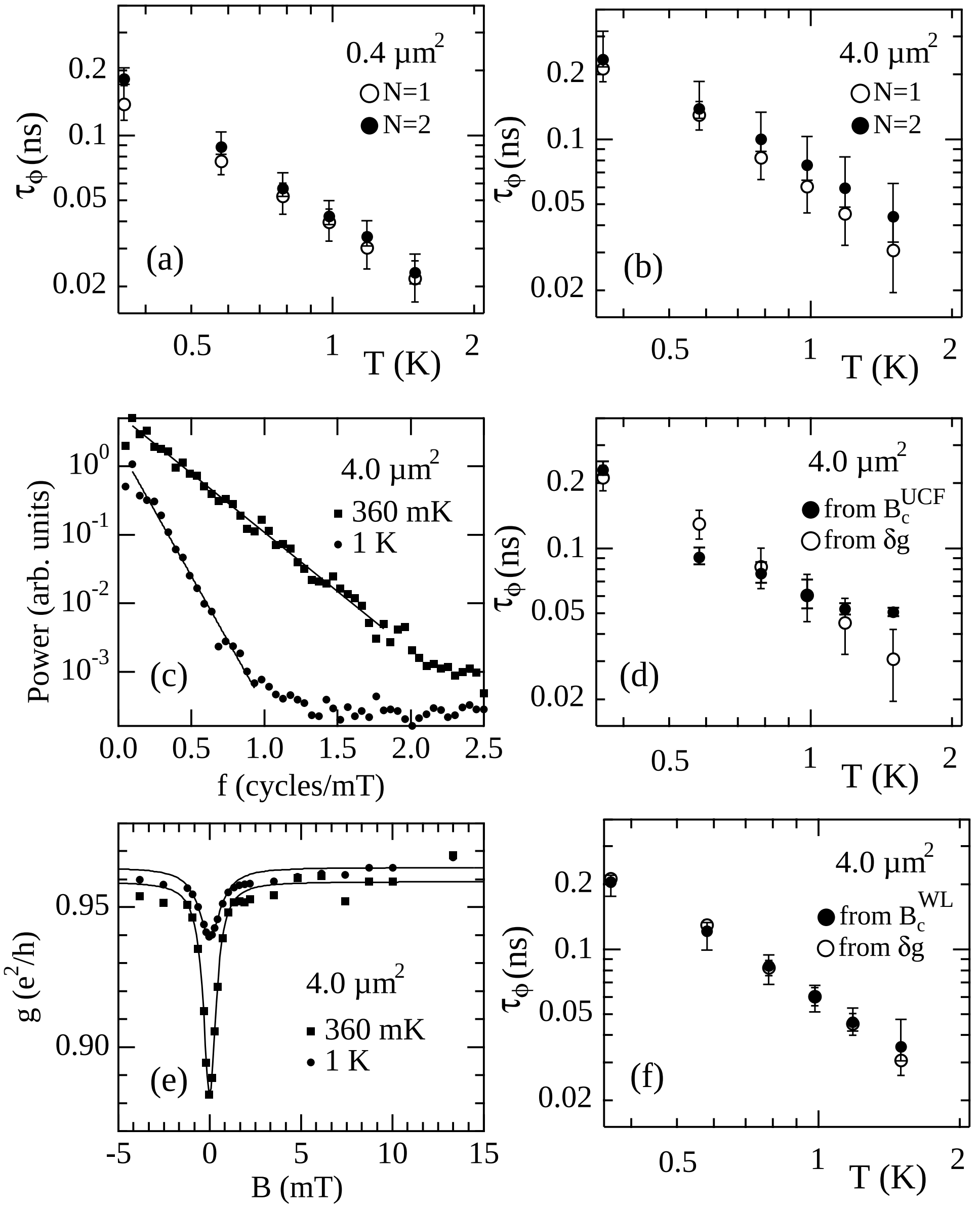

**Fig. 4.** (a), (b) Comparison of $\tau_\phi$ extracted from average $N$=1 and $N$=2 weak localization amplitude for 0.4 μm² and 4.0 μm² dots. (c) Fit of Eq. (5) to power spectral density for $N$=1 conductance fluctuations and (d) comparison of $\tau_\phi$ extracted from the characteristic field scale of UCF and from weak localization amplitude for 4.0 μm² dot. (e) Lorentzian fit (Eq. 7) to average $N$=1 weak localization lineshape and (f) comparison of $\tau_\phi$ extracted from the weak localization width of the fit, and from weak localization amplitude for 4.0 μm² dot.